# Studies on the origin of the interfacial superconductivity of Sb$_2$Te$_3$/Fe$_{1+y}$Te heterostructures

Jing Liang [1,2]†, Yu Jun Zhang [3,4]†, Xiong Yao [1], Hui Li [1], Zi-Xiang Li [5,6], Jiannong Wang [1,2], Yuanzhen Chen [4,7], and Iam Keong Sou [1,2]*

[1] Department of Physics, the Hong Kong University of Science and Technology, Hong Kong 999077, China

[2] William Mong Institute of Nano Science and Technology, the Hong Kong University of Science and Technology, Hong Kong 999077, China

[3] School of Advanced Materials, Shenzhen Graduate School Peking University, Shenzhen 518055, China

[4] Department of Physics, Southern University of Science and Technology, Shenzhen 518055, China

[5] Department of Physics, University of California, Berkeley, CA 94720, USA.

[6] Materials Sciences Division, Lawrence Berkeley National Laboratory, Berkeley, CA 94720, USA.

[7] Institute for Quantum Science and Engineering, Southern University of Science and Technology, Shenzhen 518055, China

* Correspondence and requests for materials should be addressed to I.K.S. (email: phiksou@ust.hk).

† These authors contributed equally to this work.

## Abstract

The recent discovery of the interfacial superconductivity (SC) of the $Bi_2Te_3/Fe_{1+y}Te$ heterostructure has attracted extensive studies due to its potential as a novel platform for trapping and controlling Majorana fermions. Here we present studies of another topological insulator (TI)/$Fe_{1+y}Te$ heterostructure, $Sb_2Te_3/Fe_{1+y}Te$, which also enjoys an interfacial two-dimensional SC. The results of transport measurements support that the reduction of excess Fe concentration of the $Fe_{1+y}Te$ layer not only increases the fluctuation of its antiferromagnetic (AFM) order but also enhances the quality of the SC of this heterostructure system. On the other hand, the interfacial SC of this heterostructure was found to have a wider-ranging TI-layer thickness dependence than that of the $Bi_2Te_3/Fe_{1+y}Te$ heterostructure, which is believed to be attributed to the much higher bulk conductivity of $Sb_2Te_3$ that enhances indirect coupling between its top and bottom topological surface states (TSSs). Our results provide the evidence of the interplay among the AFM order, itinerant carries from the TSSs and the induced interfacial SC of the TI/$Fe_{1+y}Te$ heterostructure system.

Three-dimensional topological insulators (3D TIs) with extraordinary electronic states have attracted intense research interest in recent years due to their unusual electronic band structures[1-4]. In an ideal 3D TI, bulk states feature an energy gap like an ordinary insulator, while surface states are characterized by a linear Dirac-like dispersion energy band with spin texture locked helically to momentum, resulting in metallic topological surface states (TSSs). The most studied 3D TIs include $Bi_2Se_3$, $Bi_2Te_3$ and $Sb_2Te_3$ binary compounds that have rhombohedral structure in which the unit cell consists of three quintuple layers (QL) bonded by van der Waals (vdW) force. Heterostructures containing a 3D TI are extensively studied because they are expected to produce novel phenomena that may not arise from heterostructures composed of conventional materials. For example, topological superconductivity (SC) has been theoretically predicted to be induced on the surface of TI-superconductor heterostructure via proximity effect, which may provide a platform to host non-Abelian Majorana fermions[5]. Interestingly, our group recently discovered a novel two-dimensional (2D) SC at the interface of two non-superconducting materials[6,7], $Fe_{1+y}Te$, a parent compound of iron-based superconductors, and $Bi_2Te_3$, a 3D TI, and later we reported further studies on this heterostructure such as the superconducting proximity effect[8], current induced depairing[9], anisotropic magnetic responses[10] and vortex dynamics[11,12]. Due to the complexity of this heterostructure, revealing the underlying mechanism of its induced 2D SC with a theoretical approach is very unlikely. Recently, experimental studies using angle-resolved photoemission spectroscopy (ARPRES)[13] and ultraviolet photoemission spectroscopy (UPS)[14] reported that the $Fe_{1+y}Te$ layer in this heterostructure involved hole doping through a transfer of electrons into the $Bi_2Te_3$ layer. However, the confirmation of this claim requires further studies to provide evidence about the linkage between the charge transfer and the behavior of the observed 2D SC.

The 2D SC of the $Bi_2Te_3/Fe_{1+y}Te$ heterostructure is likely induced on the $Fe_{1+y}Te$ layer as described in several previous works[8,14]. The studies on how some physical parameters involved in this heterostructure could modify the physical property of the $Fe_{1+y}Te$ layer and its 2D SC may provide hints to achieve a better understanding on the underlying mechanism of the induced SC. One of the most interesting issues in iron-based superconductors is the interplay between magnetism and SC. A synthetized bulk $Fe_{1+y}Te$ crystal usually favors excess Fe atoms in interstitial sites, which could significantly affect its magnetic order and phase transitions[15-17]. Below 70K, for bulk $Fe_{1+y}Te$ with $0.04 < y < 0.11$, the spin direction of Fe atoms is oriented along the diagonal of the Fe-Fe square lattice, forming a bi-collinear commensurate antiferromagnetic (AFM) structure, followed by a mixed phase for $0.11 < y < 0.13$ and then an incommensurate AFM magnetic order for $0.13 < y < 0.15$ as $y$ increases[16]. It was observed that the commensurate AFM transition temperature for bulk $Fe_{1+y}Te$ increases as the concentration of excess Fe decreases down to $y = 0.04$. Studies on superconducting iron chalcogenide ternary $Fe_{1.02}Te_{1-x}Se_x$[18] and iron pnictides such as $ReFeAsO_{1-x}F_x$ (Re = rare earth element)[19] showed that AFM order is antagonistic to SC that generally occurs after the magnetic order was suppressed by varying chemical doping concentration, even though they can coexist in a certain composition range[20,21].

In this study, we demonstrate that 2D SC also appears in another $TI/Fe_{1+y}Te$ system, the $Sb_2Te_3/Fe_{1+y}Te$ heterostructure. We have systematically studied how the quality of the 2D SC of this heterostructure depends on the fluctuation of the AFM order of the $Fe_{1+y}Te$ layer and the thickness of the $Sb_2Te_3$ layer. Particularly, the results from the former study provide a deeper understanding on the correlation between SC and AFM in this heterostructure while the results from the latter study support that the TSSs may be involved in the induction of SC. Based on these experimental observations, we propose a possible origin of the induced interfacial SC.

## Results

**Structural characterizations of the $Sb_2Te_3$/$Fe_{1+y}Te$ heterostructure.** Reflection high-energy electron diffraction (RHEED) patterns were captured after the growth of $Fe_{1+y}Te$ and $Sb_2Te_3$ layers. Figure 1a, b displays the RHEED patterns of tetragonal $Fe_{1+y}Te$ when the incident electron beam is along the $Fe_{1+y}Te$ [100] and [1$\bar{1}$0] axis with the sample rotation angle at $\varphi=0°$ and $\varphi=45°$, respectively. The diffraction streaks are marked by short lines and denoted by 2D Miller indices. The streaky patterns of $Fe_{1+y}Te$ display a 90° rotational symmetry, as expected from the tetragonal structure of $Fe_{1+y}Te$. In contrast, the RHEED pattern of $Sb_2Te_3$ at rotation angle $\varphi=0°$ shown in Fig. 1c has two sets of streaks. The spacing ratio of the wider and narrower set of streaks is approximately $\sqrt{3}$, and their 2D Miller indices are marked correspondingly. As the sample rotates, the diffraction pattern of $Sb_2Te_3$ repeats at every 30°, which can be explained by the fact that the as-grown $Sb_2Te_3$ layer consists of two hexagonal lattices twisted by 90°. This is attributed to the four-fold symmetry of the bottom $Fe_{1+y}Te$ lattice and the six-fold symmetry of the top $Sb_2Te_3$ lattice with respect to the growth direction. The details will be addressed in next paragraph where the results of the cross-sectional scanning transmission electron microscopy (STEM) imaging are presented.

The epitaxial growth of bismuth and antimony chalcogenides on substrates that have six-fold symmetry, such as $Al_2O_3$(0001)[22], Si(111)[23], $BaF_2$(111)[24,25] InP(111)[26] and GaAs(111)[27], usually results in two mirror-symmetric twin domains due to the fact that there are two stacking sequences (ABCAB and ACBAC) along the [0001] direction for one QL of these chalcogenides.

In order to investigate whether such twin domains exist in $Sb_2Te_3$ grown on $Fe_{1+y}Te$ that has four-fold symmetry and how they are developed into a different symmetry observed by RHEED as described above, high-resolution spherical-aberration-corrected STEM imaging was performed on a $Sb_2Te_3$(27 QL)/$Fe_{1+y}Te$(60 nm) heterostructure (1 QL of $Sb_2Te_3$ is approximately equal to 10.2 Å), which was expected to provide more visualized and finer structural analysis at atomic scale than what can be achieved by RHEED. Cross-sectional high-angle annular dark field (HAADF) STEM images of this heterostructure observed at different locations of one TEM specimen with the zone axis along the [100] direction of $Fe_{1+y}Te$ are displayed in Fig. 1d-f. As can be seen, Fig. 1d, e displays the atomic arrangement of Te-Sb-Te-Sb-Te QL stacking in ABCAB and ACBCB sequence with the [11$\bar{2}$0] direction (Fig. 1d) and [$\bar{1}\bar{1}$20] direction of $Sb_2Te_3$ (Fig. 1e) aligned with the [100] direction of $Fe_{1+y}Te$, respectively. In Fig. 1g, h, schematic drawings with both top-view and side-view modes for these two stacking sequences are displayed. As mentioned earlier, the $Sb_2Te_3$ layer was grown on $Fe_{1+y}Te$[001] that has a four-fold symmetry, thus it is expected that one can find two other domains that are laterally twisted by 90° from each of the two domains schematically drawn in Fig. 1g, h. Figure 1i, j shows the corresponding drawings of these expected domains. One can see that these two domains have identical side-view atomic arrangement when the viewing zone axis is along the [100] direction of $Fe_{1+y}Te$. In fact, we did observe such an atomic arrangement among our STEM images and one typical example is shown in Fig. 1f. These observations indeed provide evidence that the $Sb_2Te_3$ layer consists of two hexagonal lattices twisted by 90°. It is worth to point out that the four different domains mentioned above can be generated through azimuthal rotation of a domain having a stacking sequence either shown in Fig. 1g or h, then laterally twisted by 90°, 180° and 270°. This can be visualized by starting with the domain in Fig. 1g, then Fig. 1i, h and j are the resulted domains when it is twisted by 90°, 180° and 270°, respectively. It is also worth mentioning that all the STEM-HAADF images shown in Fig. 1d-f enjoy a sharp and defect-free interface, which can be attributed to the vdW bonding nature between $Sb_2Te_3$ and $Fe_{1+y}Te$. Based on these images, the lattice parameters of the $Fe_{1+y}Te$ unit cell at the interface are determined to be $a = 3.88$ Å and $c = 6.24$ Å, and those of $Sb_2Te_3$ are $a = 4.24$ Å and $c = 30.54$ Å, which are close to their reported bulk values. As shown in Supplementary Fig. 1, high resolution X-ray diffraction studies support that all the four domains of $Sb_2Te_3$ likely orient with their $c$-axes along the growth direction.

**Two-dimensional superconductivity.** Figure 2a, b shows the temperature dependent resistance of the $Sb_2Te_3$(27 QL)/$Fe_{1+y}Te$ heterostructure measured under different magnetic fields in directions perpendicular ($H_\perp$) and parallel ($H_{//}$) to the interface, respectively. At zero applied magnetic field, resistance starts to drop at an onset temperature $T_C^{onset}$ of 12.3 K and zero-resistance state happens at 3.1 K. Figure 2a, b displays a large anisotropy regarding the direction of the applied magnetic field since the transitions significantly broaden as $H_\perp$ increases, while such a broadening is much weaker as $H_{//}$ increases to the same magnitudes, indicating the SC is likely in 2D. Its further confirmation could be obtained by studying the temperature dependence of the upper critical magnetic field $\mu_0 H_{c2}$ of the observed SC. Here, we define the critical superconducting transition temperature $T_C$ as the resistance drops to 50% of the normal state value at 15 K. The out-of-plane upper critical magnetic field and in-plane upper critical field are denoted as $\mu_0 H_{c2}^\perp$ and $\mu_0 H_{c2}^{//}$ respectively. Figure 2c shows the temperature dependence of the upper critical field $\mu_0 H_{C2}(T)$ for the $Sb_2Te_3$(27 QL)/$Fe_{1+y}Te$ heterostructure in directions parallel ($H_{//}$) and perpendicular ($H_\perp$) to the interface, where the $x$ axis denotes the $T_C$ values at different applied magnetic field. It is well known that 2D SC is governed by Ginzburg-Landau (GL) theory[28] with the temperature dependences of the upper critical magnetic fields expressed by

$$\mu_0 H_{c2}^{\perp} = \frac{\Phi_0}{2\pi \xi_{GL}^2(0)} \left(1 - \frac{T}{T_C(0)}\right) \quad (1)$$

$$\mu_0 H_{c2}^{//} = \frac{\Phi_0 \sqrt{12}}{2\pi \xi_{GL}(0) d_{SC}} (1 - \frac{T}{T_C(0)})^{\frac{1}{2}} \quad (2)$$

where $\Phi_0$ is the magnetic flux quantum, $\xi_{GL}(0)$ the zero-temperature GL in-plane coherence length, $d_{SC}$ temperature-independent SC thickness, and $T_C(0)$ the critical temperature at zero magnetic field. Our data shown in Fig. 2c indeed can be well fitted by this theory in both parallel and perpendicular directions, the solid curves in this figure are resulted from the best fitting with $\xi_{GL}(0) = 4.1 \pm 0.2$ nm and $d_{SC} = 3.9 \pm 0.2$ nm. It is important to point out that neither a pure $Fe_{1+y}Te$ nor a pure $Sb_2Te_3$ thin film grown under similar conditions exhibits SC (see Supplementary Fig. 2), further supporting that the observed SC in the $Sb_2Te_3/Fe_{1+y}Te$ heterostructure occurs at its interface.

It is commonly accepted that the transport properties of a 2D SC system can be well described by the 2D Berezinskii-Kosterlitz-Thouless (BKT) theoretical model[29-31], in which vortex-antivortex pairs are formed and bound together below a critical temperature called BKT temperature $T_{BKT}$. At temperature above $T_{BKT}$, vortex-antivortex pairs are thermally dissociated and free vortices give rise to a finite resistance. At temperature just above $T_{BKT}$, the temperature-dependent resistance at zero magnetic field is predicted to be in a form of

$$R = R_0 \exp(-bt^{-1/2}) \quad (3)$$

where $R_0$ and $b$ are material-specific parameters and $t = \frac{T}{T_{BKT}} - 1$ is the reduced temperature. The main panel of Fig. 2d shows the best fit using Eq. (3) for the temperature dependent resistance at zero applied magnetic field of the heterostructure, yielding $T_{BKT} = 3.68$ K. The insert of Fig. 2d shows the extrapolation of the linear part of $(d\ln R/dT)^{-2/3}$ to zero, yielding a consistent value of $T_{\text{BKT}} = 3.64$ K. Another important piece of evidence of BKT transition is obtained from fitting the voltage-current (*V-I*) curves on a log-log scale for a temperature range from 2 to 12 K near each critical current in its superconducting transition region with a $V \propto I^{\alpha(T)}$ power-law dependence, where α equals 3 at $T_{BKT}$, as shown in Fig. 2e. Figure 2f plots α as a function of temperature from which $T_{BKT}$ is determined to be 3.71 K. The above analyses thus provide strong evidence for the 2D nature of the observed SC of the $Sb_2Te_3/Fe_{1+y}Te$ heterostructure.

**The role of fluctuation of the antiferromagnetic order.** Aiming at studying the interplay between the AFM order and SC, a group of $Sb_2Te_3$(24 QL)/$Fe_{1+y}Te$(60 nm) heterostructures named SF-1, 2 and 3 and two pure $Fe_{1+y}Te$(60 nm) thin films named F-1 and 2 were fabricated with various excess Fe concentrations. A combination of X-ray photoelectron spectroscopy (XPS) and spherical-aberration-corrected STEM was used to estimate the nominal *y* values of these samples. For achieving a more accurate analysis, a standard sample with known chemical composition is needed to derive the more reliable values of the relative sensitivity factors (RSFs) used in XPS analysis. These RSFs are further assumed to be constants for a certain variation of the sample matrix. We used sample SF-1 as such a standard sample and its *y* value was estimated by counting the number of interstitial Fe atoms in its spherical-aberration-corrected STEM images of its $Fe_{1+y}Te$ layer. Figure 3 shows one of the STEM-HAADF images that have the highest detected amount of interstitial Fe, in which the rectangular boxes contain the interstitial Fe atoms. Since fewer interstitial Fe atoms are seen in some of these images, we obtained the estimated *y* value of this sample as 0.004 by taking the average. The nominal *y* values of SF-2,

SF-3, F-1 and F-2, as estimated using the peak areas of Fe and Te in their XPS spectra and the RSFs derived from the estimated $y$ value of sample SF-1, are 0.021, 0.074, 0.042 and 0.09, respectively.

The normalized resistances as a function of temperature for sample F-1 and 2 are shown in Fig. 4a. Sample F-1 with fewer excess Fe concentration of the two samples shows a semiconducting behavior within the entire measured temperature range from 300 K to 2 K. Sample F-2 with more excess Fe shows a semiconducting to metallic transition at 47.3 K, which is known to be associated with the AFM phase transition. The spin fluctuation of the AFM order in Fe-SCs was previously proposed to be corelated to their SC[32,33]. We believe that such fluctuation can be reflected by the distinctness of the AFM transition, that is, less distinctive this transition corresponds to higher spin fluctuation, which is consistent with the observations on the temperature-dependent resistivity curves of $Fe_{1.02}(Te_{1-x}Se_x)$ alloys[18]. Interstitial Fe atoms are believed to provide magnetic proximity coupling with the neighboring AFM spins (as we addressed in a previous study[8]) to reduce their fluctuation. The results shown in Fig. 4 indicate that the reduction of excess Fe concentration in $Fe_{1+y}Te$ layer increases the fluctuation of this magnetic ordering, consistent with the previous findings from Bao et al.'s studies using a spin polarized inelastic neutron spectrometer[17]. Figure 4b shows the temperature-dependent normalized resistances of samples SF-1, 2 and 3. Sample SF-1 and SF-2 with less excess Fe than sample SF-3 show a semiconducting behavior as the temperature decreases from 300 K to just above the superconducting transition temperature, while SF-3 shows an AFM transition at 47.6 K. This correlation is consistent with that observed for the two pure $Fe_{1+y}Te$ layers as shown in Fig. 4a. As can been seen in Fig. 4b, the onset temperatures of superconducting transition are 9.2 K, 8.9 K and 8.7 K for SF-1, 2 and 3, respectively. These results indicate that the reduction of excess Fe concentration not only increases the spin fluctuation of the AFM order in the $Fe_{1+y}Te$ layer but also enhances the quality of the SC of this heterostructure system. In fact, this provides important evidence that spin fluctuation of the AFM order in the $F_{1+y}Te$ layer of the heterostructure is related to the induction of the observed 2D SC.

**$Sb_2Te_3$ thickness dependence on the interfacial superconductivity.** In the following, we will present the studies on the $Sb_2Te_3$ layer thickness dependence of the 2D SC of the $Sb_2Te_3/Fe_{1+y}Te$ heterostructures with the aim at providing the evidence that the TSSs of the $Sb_2Te_3$ layer are likely involved in the induction of the SC. A set of $Sb_2Te_3/Fe_{1+y}Te$(60 nm) heterostructures, each containing a $Sb_2Te_3$ layer with a different thickness, was fabricated for these studies. Figure 5a shows the normalized temperature dependent resistance $R/R(T=100 K)$ curves of $Sb_2Te_3/Fe_{1+y}Te$ heterostructures with $Sb_2Te_3$ of 10, 15, 24 and 27 QL, which display SC with onset temperatures of 3.7, 6.6, 8.7 and 12.3 K, respectively, showing a trend that the onset temperature of SC increases with the $Sb_2Te_3$ thickness. For better illustrating the SC signature of the $Sb_2Te_3$(10 QL)/$Fe_{1+y}Te$ sample with the thinnest $Sb_2Te_3$ layer among the four heterostructures, the insert in Fig. 5a shows the resistance versus temperature curve for this sample near the SC transition, which indeed displays a resistance drop at 3.7 K. Figure 5b shows the zoomed in normalized resistance curves for these heterostructure samples around their AFM transition temperatures, which are 48.2, 53.3, 47.6 and 53.3 K for $Sb_2Te_3$ thickness of 10, 15, 24 and 27 QL, respectively. The variation of the AFM transition temperatures of these heterostructures is caused by the small differences in the spin fluctuation of the AFM order of their $Fe_{1+y}Te$ layers. As can be seen in Fig. 5b, $Sb_2Te_3/Fe_{1+y}Te$ samples with 10 QL- and 24 QL-thick $Sb_2Te_3$ have very similar AFM transition temperatures, which indicates that the spin fluctuation in these two samples are nearly the same, however, the latter has a much higher onset SC temperature than the former. A similar circumstance also exists for the other two $Sb_2Te_3/Fe_{1+y}Te$ heterostructures with 15 QL and 27 QL of $Sb_2Te_3$. Thus, the $Sb_2Te_3$ thickness dependence of the onset SC temperature revealed in Fig. 5a is a true property of these heterostructures. In Supplementary Information, we described an approach to fabricate two heterostructures with $Sb_2Te_3$ layers of different thicknesses grown on a single $Fe_{1+y}Te$ layer with a high spin fluctuation. The results of their SC characteristics as shown in Supplementary Fig. 3 indeed provide further confirmation of this $Sb_2Te_3$ thickness dependence.

**Spin-orbit coupling alone cannot induce the observed superconductivity.** In the following, we will briefly address the studies of another factor that was thought to be the underlying cause of the 2D SC in the TI/$Fe_{1+y}$Te heterostructure system. It is related to an obvious and straightforward theoretical hypothesis that the strong SOC of TI materials could itself induce the 2D SC. One can test this hypothesis experimentally by studying the transport properties of a $Fe_{1+y}$Te heterostructure with a top layer possessing strong SOC but it is not a TI. We chose elemental Sb and Bi thin films as the replacement for the TI layers for the following reasons: The difference between $A_2B_3$-type TI compounds and Sb is that the former compounds have a bulk energy gap with TSSs confined at the surfaces, while the latter is a semimetal[35,36] and thus its TSSs penetrate to the entire interior of the film[37]. On the other hand, due to an even stronger SOC in Bi (which is stronger than that of both $Bi_2Te_3$ and $Sb_2Te_3$), its surface bands inverse twice along $\bar{\Gamma}\bar{M}$ and return to the valence band, making Bi a semimetal with topologically trivial surface states[38] (except that ultra-thin Bi films are a 2D TI[39]). Thus, it is expected that both Sb/$Fe_{1+y}$Te and Bi/$Fe_{1+y}$Te heterostructures can be used to test if SOC alone could induce interfacial SC in $Fe_{1+y}Te$. Supplementary Fig. 4 displays the temperature dependent resistance curves of three heterostructures, $Sb(10\ nm)/Fe_{1+y}Te$, $Bi(13\ nm)/Fe_{1+y}Te$ and $Sb_2Te_3(18\ QL)/Fe_{1+y}Te$. As shown in this figure, the AFM transition temperatures for these three heterostructures are 54.8, 55.8 and 61.3 K, respectively, indicating that the AFM order of the $Sb_2Te_3/Fe_{1+y}Te$ heterostructure is stronger (equivalent to weaker spin fluctuation) than that of both the $Sb/Fe_{1+y}Te$ and $Bi/Fe_{1+y}Te$ heterostructures. As can be seen in Supplementary Fig. 4, the $Sb_2Te_3/Fe_{1+y}Te$ heterostructure is superconducting as expected, however, neither the $Sb/Fe_{1+y}Te$ nor the $Bi/Fe_{1+y}Te$ heterostructure displays SC and they only show an up-turn in resistance at low temperature instead. These results indicate that SOC alone is not able to induce the interfacial SC observed in $TI/Fe_{1+y}Te$ heterostructures.

## Discussions

We believe that the observed $Sb_2Te_3$-thickness dependence on the SC quality of the heterostructure system described earlier can be linked to the evolution of the TSSs at the interface in the following ways. The degree of integrity (totality) of the TSSs at the interface increases with the $Sb_2Te_3$ thickness, which likely corresponds to higher SC quality of the heterostructure system. For thin $Sb_2Te_3$ layers, the two factors that can affect the degree of integrity of the interfacial TSSs are direct coupling and indirect coupling between the top and bottom TSSs. In direct coupling, the wave functions of the top and bottom TSSs are overlapped such that a small gap opens at the Dirac point, which reduces the integrity of the bottom TSSs. For $Sb_2Te_3$, the critical thickness for observing gapless TSSs is 4 QL as demonstrated by scanning tunneling microscope (STM)[40,41] and ARPES[42]. Since the observed $Sb_2Te_3$ thickness dependence in this study covers a thickness range from 10 to 27 QL, direct coupling should not play a key role in this dependence. Jiang et al. previously reported that molecular beam epitaxy (MBE)-grown $Sb_2Te_3$ usually displays a p-type conductivity due to various intrinsic defects[43]. We have also confirmed the p-type nature of our $Sb_2Te_3$ layers by conducting Hall measurements (see Supplementary Fig. 2). This bulk conductivity was known to mediate a so-called indirect coupling between the top and bottom TSSs[44], which likely occurs even when the $Sb_2Te_3$-thickness is larger than the critical thickness of direct coupling. We believe this indirect coupling can also reduce the integrity of the TSSs at the interface of the heterostructure and the coupling is weakened as the TI thickness increases[45]. Thus, indirect coupling is considered to be mainly responsible for the observed $Sb_2Te_3$-thickness dependence on the 2D SC of the heterostructure system. It is worth pointing out that such a TI-layer thickness dependence beyond the critical thickness for observing gapless TSSs of the $Bi_2Te_3/Fe_{1+y}Te$ heterostructures is less obvious as shown in our previous work[7]. This can be explained by the big contrast between the estimated bulk carrier concentration of a $Sb_2Te_3$(25 QL) and that of a $Bi_2Te_3$(50 QL) thin film, the former is $1.63 \times 10^{22}$ $cm^{-3}$ as given in Supplementary Fig. 2 of this work, while the latter is about $1.3$-$1.6 \times 10^{19}$ $cm^{-3}$ as reported by us earlier[46]. This big contrast implies that our $Sb_2Te_3$ thin film has a much higher bulk conductivity than $Bi_2Te_3$, leading to a stronger indirect coupling between the top and bottom TSSs for $Sb_2Te_3$ and in consequence a wider-ranging TI-thickness dependence of the 2D SC as described above.

The fact that 2D SC of a $Fe_{1+y}Te$ layer occurs only when it is covered with either $Bi_2Te_3$ or $Sb_2Te_3$ implies that a property of TI is likely involved in the induction of this SC. Since the TI layer in these heterostructures provides a source of itinerant carriers confined at the interface through the TSSs, it is possible that these itinerant carriers exhibit an unconventional pairing at the interface, leading to the observed interfacial SC. Recently, Yasuda et al. observed a large nonreciprocal charge transport associated with the superconducting transition of $Bi_2Te_3$/FeTe, which is revealed to originate from the current-induced-modulation of supercurrent density due to spin-momentum locking, indeed suggesting a close connection between SC and TSSs[34].

Below we briefly discuss the possible pairing mechanism of the 2D SC at the $Sb_2Te_3$/$Fe_{1+y}$Te interface. We believe the two key ingredients at the $Sb_2Te_3$/$Fe_{1+y}$Te interface are (a) 2D itinerant electrons induced by the TSSs of $Sb_2Te_3$, and (b) local spin moments with AFM ordering in $Fe_{1+y}$Te. The hybridization of the two ingredients, which are reminiscent to the Anderson lattice, was recently proposed to explain the strong SC pairing in strongly correlated materials where SC and magnetism are intertwined, for instance electron-doped FeSe[33]. At the $Sb_2Te_3$/$Fe_{1+y}$Te interface, the fluctuation of the local moments of the AFM ordering in $Fe_{1+y}$Te is enhanced due to the hybridization with the itinerant electrons in TSSs. On the other hand, the enhanced fluctuating spin moments provide a driving force for the SC pairing of the itinerant electrons at the interface. The fact that SC transition temperature increases upon the increase of the fluctuation of the AFM order, as observed in this study, also implies that SC pairing benefits from the fluctuation of AFM order. Consequently, we propose that the Cooper pairing mediated by spin fluctuation enhanced by the hybridization of local spin moments in $Fe_{1+y}$Te and 2D itinerant electrons in the TSSs of $Sb_2Te_3$ is a possible mechanism of the SC observed in the $Sb_2Te_3$/$Fe_{1+y}$Te heterostructures. More quantitative experimental analysis and theoretical calculation are desired to confirm such mechanism in future studies. We believe that the $Sb_2Te_3$/$Fe_{1+y}$Te heterostructures shall provide a new platform not only for investigating the novel physics emerging from the interplay among SC, magnetism and topological helical surface states but also for trapping and controlling Majorana fermions.

## Method

**Sample preparation.** All the samples studied in this work were fabricated by a VG-V80H MBE system equipped with *in situ* RHEED. Sample synthesis was conducted using high purity $Sb_2Te_3$ and ZnSe compound sources together with Fe and Te elemental sources. A ZnSe buffer layer (80 nm) was firstly deposited on the GaAs (001) n+ substrates to form a flat surface, followed by a deposition of $Fe_{1+y}Te$ with a thickness of 60 nm and a $Sb_2Te_3$ layer. For the studies of the interstitial-Fe concentration dependence, a set of $Sb_2Te_3$(24 QL)/$Fe_{1+y}Te$ heterostructures and a set of pure $Fe_{1+y}Te$ thin films with various amount of interstitial Fe in the $Fe_{1+y}Te$ layer were fabricated by changing the temperature of Te effusion cell while keeping the temperatures of the Fe cell and substrate constant. Another set of $Sb_2Te_3$/$Fe_{1+y}Te$ heterostructure samples were fabricated containing $Sb_2Te_3$ thin films with average thicknesses of 10, 15, 24 and 27 QL to study the $Sb_2Te_3$-thickness dependence of the 2D SC. All the samples were capped with a Te protective layer (~8 nm) to avoid oxidization.

**Characterizations.** The thickness of each layer was determined by TEM, JEOL 2010F. High resolution spherical-aberration-corrected STEM images were carried out in a JEM-ARM200F TEM equipped with a probe corrector and a HAADF detector for finer crystal structure analysis. $Sb_2Te_3$/$Fe_{1+y}Te$ heterosturctures were cut into long strips (with dimension∼2 mm × 6 mm) and conventional four-point electric contacts were made on the surface of each strip using silver paint as the contact material for performing *ex situ* electrical and magneto-transport measurements, which were carried out in a physical property measurement system equipped with a 14 T superconducting magnet. To determine the interstitial Fe concentration in the $Fe_{1+y}Te$ layer, *ex situ* XPS measurements were performed by Kratos-Axis Ultra DLD XPS equipped with a monochromatic Al kα x-ray source (photon energy 1486.7 eV, 150 W), which was operated in hybrid lens mode with an energy step of 100.0 meV, a pass energy of 40 eV and a large measuring area of $1 \times 2$ mm$^2$. Before acquiring a spectrum, Ar ion sputtering was performed to remove the upper layer(s) to reveal a pure surface of the Fe$_{1+y}$Te layer. The binding energy was calibrated with the C 1s peak at 284.8 eV. The actual core-level peak areas were determined after applying Shirley background subtraction. HRXRD measurements were conducted using a PANalytical multipurpose diffractometer with an X'celerator detector (PANalytical X'Pert Pro) for crystalline characterizations.

**Acknowledgments:**

The authors acknowledge support from the Materials Characterization and Preparation Facility, the Hong Kong University of Science and Technology, for providing most of the chemical and structural characterizations, including the aberration-corrected TEM characterization facility under the CRF project (project number: C6021-14E). This research was funded by the Research Grants Council of the Hong Kong Special Administrative Region, China, grant number 16304515, 16301418 and C6013-16E, and William Mong Institute of Nano Science and Technology, project number WMINST19SC07.

**Author Contributions**

J.L. and I.K.S. initiated this study and further designed the experiments with contributions from other authors; J.L. carried out the sample growth and structural characterization with contributions from X.Y.; Y.J.Z contributed the transport measurements with contributions from X.Y., and H.L.; J.L., Y.J.Z, X.Y., H.L., Z.X.L. and I.K.S. performed the data analysis; All authors contributed to the discussions; I.K.S. and J.L. wrote the manuscript with contributions from other authors.

**Competing Financial Interests statement**

The authors declare no competing financial interests.

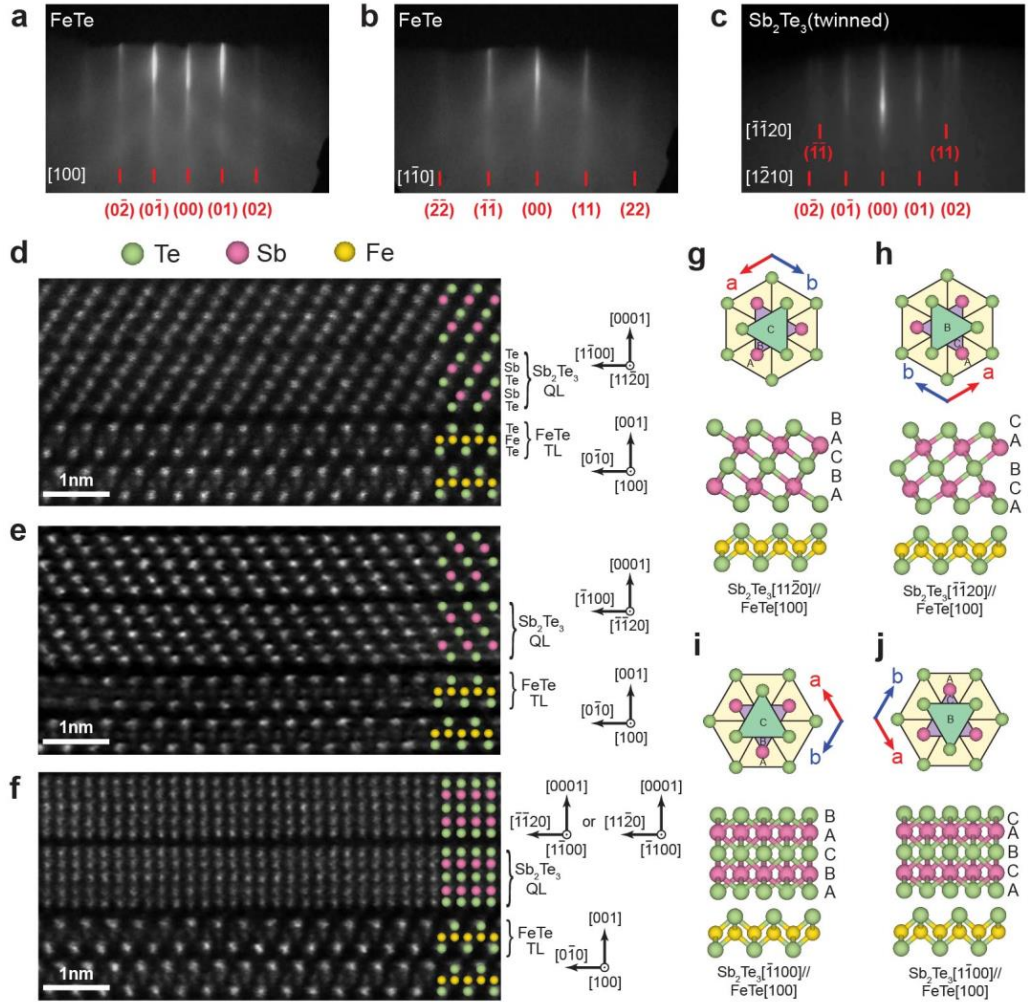

**Fig. 1** Structural analysis of a $Sb_2Te_3/Fe_{1+y}Te$ heterostructure. **a**, **b** RHEED patterns of $Fe_{1+y}Te$ taken when the electron beam is along (**a**) [100] and (**b**) [1$\bar{1}$0] direction. **c** RHEED pattern of $Sb_2Te_3$ taken when the electron beam is along the [100] direction of $Fe_{1+y}Te$. The short lines in (**a**-**c**) denote 2D Miller indices of the observed diffraction streaks. **d**-**f** Three cross-sectional STEM-HAADF images of the heterostructure taken at different locations of one TEM specimen with the zone axis along the [100] direction of $Fe_{1+y}Te$. **g**, **h** Schematic drawings of top and side views of the atomic arrangement of domain (**d**) and (**e**), respectively. **i**, **j** Schematic drawings of top and side views of the other two domains that have identical side-view atomic arrangement corresponding to the STEM-HAADF image shown in (**f**).

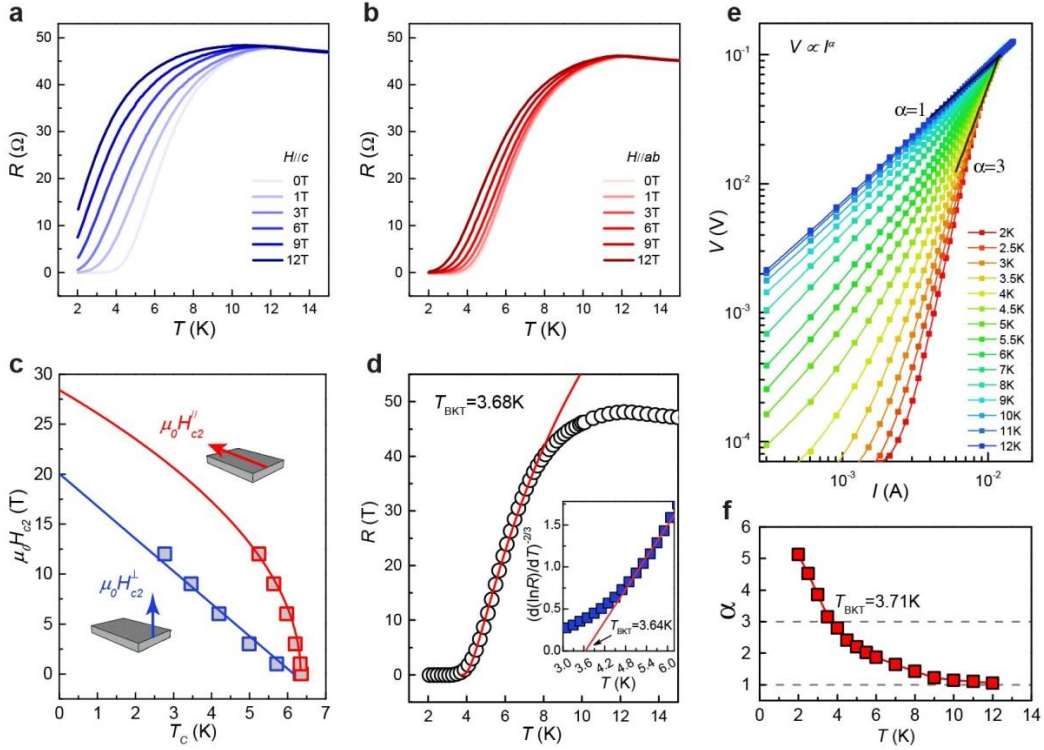

**Fig. 2** 2D SC nature of the $Sb_2Te_3$(27QL)/$Fe_{1+y}$Te heterostructure. **a, b** Resistance as a function of temperature under (**a**) out-of-plane and (**b**) in-plane magnetic field up to 12 T. **c** Temperature dependence of the upper critical field $\mu_0 H_{c2}^{\perp}$ and $\mu_0 H_{c2}^{//}$. Solid curves are obtained by fitting with the theoretical 2D Ginzburg-Landau equations. **d** Main-panel: Temperature dependence of resistance with a fitted curve based on the Berezinskii-Kosterlitz-Thouless (BKT) model, yielding $T_{BKT} = 3.68$ K. Insert: $[dlnR/dT]^{-2/3}$ as a function of temperature with its linear part extrapolated to zero, yielding $T_{BKT} = 3.64$ K. **e** V-I curves at various temperatures plotted on a logarithmic scale. The line with $\alpha = 1$ represents ohmic behavior while the line with $\alpha = 3$ corresponds to the BKT transition. **f** $\alpha$ versus $T$ extracted from fitting the data in **e** with power law $V \propto I^{\alpha}$, showing $\alpha$ approaches 3 at $T_{BKT} = 3.71$ K.

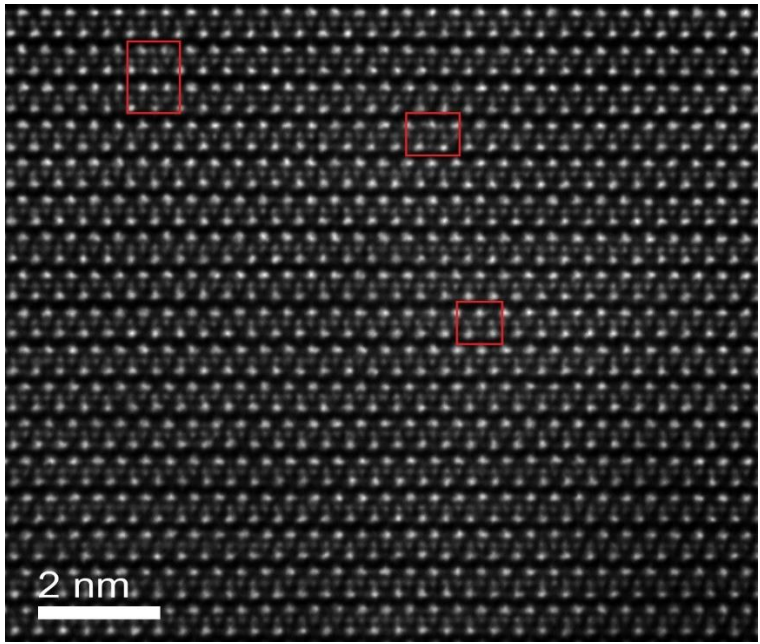

**Fig. 3** A cross-sectional STEM image of sample SF-1 taken in the $Fe_{1+y}Te$ region. The red rectangles enclose the areas that have interstitial Fe atoms.

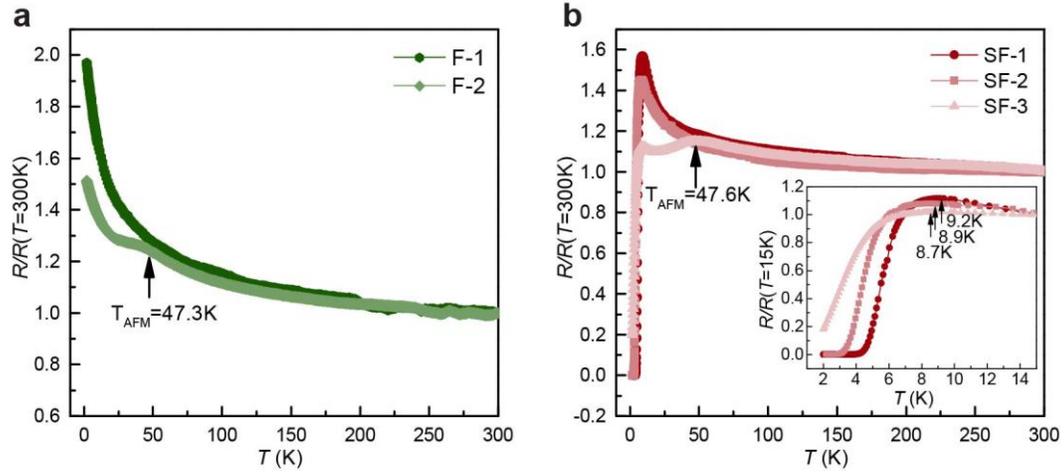

**Fig. 4** The interplay among the excess Fe concentration, magnetism and SC. **a, b** Normalized temperature-dependent resistance of (**a**) two pure $Fe_{1+y}Te$ (60 nm) films named F-1 and 2; and (**b**) a group of $Sb_2Te_3$ (24 QL)/$Fe_{1+y}Te$ (60 nm) named SF-1, 2 and 3 with various excess Fe concentration $y$. Insert of (**b**) shows a zoomed-in region at low temperature from 1 K to 15 K for displaying the contrast of the SC transition of these heterostructures.

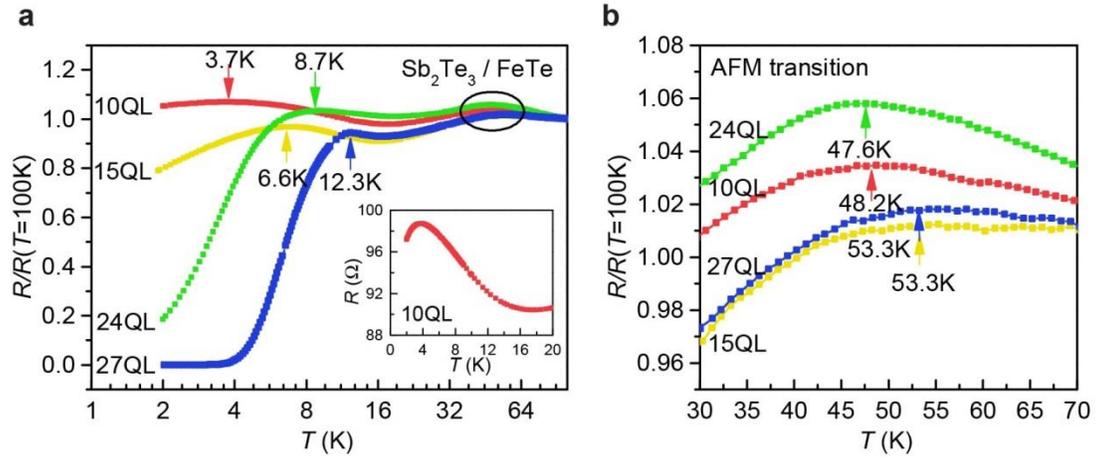

**Fig. 5** The influences of $Sb_2Te_3$ thickness on the SC of $Sb_2Te_3/Fe_{1+y}Te$ heterostructures. **a** Normalized resistance as a function of temperature for $Sb_2Te_3/Fe_{1+y}Te$ heterostructures with varying thicknesses of the $Sb_2Te_3$ thin films of 10, 15, 24 and 27 QL. The insert shows the resistance at low temperature for the $Sb_2Te_3/Fe_{1+y}Te$ sample with 10 QL $Sb_2Te_3$ layer for clarifying the signature of its SC. **b** Normalized resistance for these four heterostructures in a zoomed-in region around the antiferromagnetic (AFM) transition. The arrows mark the AFM transition temperatures.

# Supplementary Information for:

# Studies on the origin of the interfacial superconductivity of Sb$_2$Te$_3$/Fe$_{1+y}$Te heterostructures

Jing Liang [1,2]†, Yu Jun Zhang [3,4]†, Xiong Yao [1], Hui Li [1], Zi-Xiang Li [5,6], Jiannong Wang [1,2], Yuanzhen Chen [4,7], and Iam Keong Sou [1,2]∗

[1] Department of Physics, the Hong Kong University of Science and Technology, Hong Kong 999077, China

[2] William Mong Institute of Nano Science and Technology, the Hong Kong University of Science and Technology, Hong Kong 999077, China

[3] School of Advanced Materials, Shenzhen Graduate School Peking University, Shenzhen 518055, China

[4] Department of Physics, Southern University of Science and Technology, Shenzhen 518055, China

[5] Department of Physics, University of California, Berkeley, CA 94720, USA.

[6] Materials Sciences Division, Lawrence Berkeley National Laboratory, Berkeley, CA 94720, USA.

[7] Institute for Quantum Science and Engineering, Southern University of Science and Technology, Shenzhen 518055, China

∗ Correspondence and requests for materials should be addressed to I.K.S. (email: phiksou@ust.hk).

† These authors contributed equally to this work.

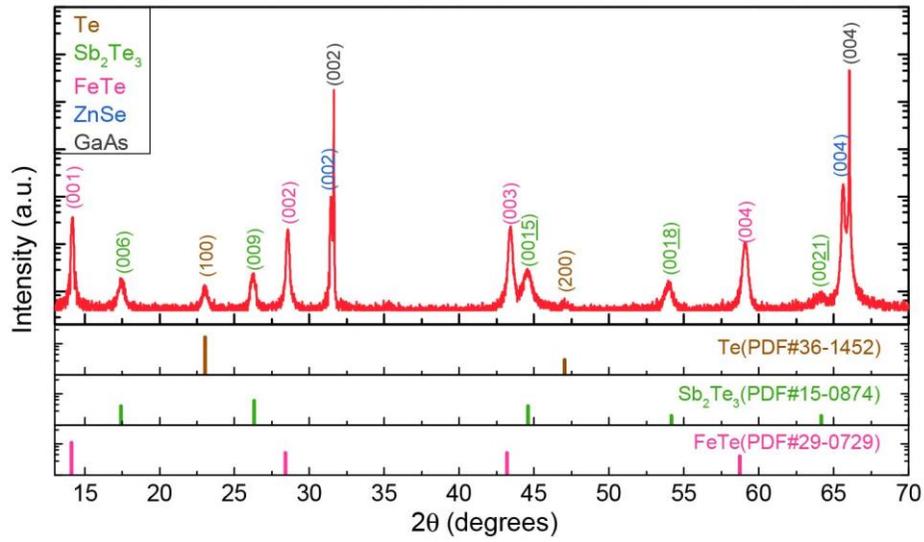

**Supplementary Figure 1** | Symmetric high resolution X-ray diffraction of the $Sb_2Te_3$(27 QL)/$Fe_{1+y}Te$(60 nm) heterostructure. Lower part shows powder diffraction files for the three compounds contained in this sample.

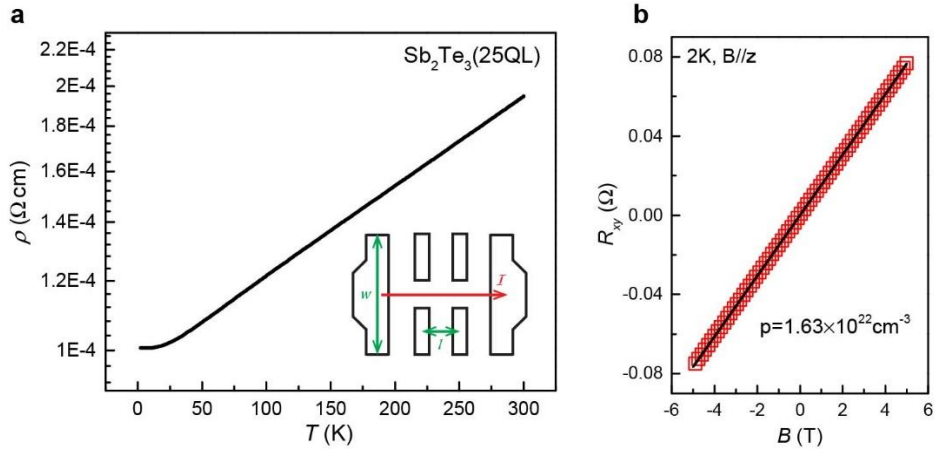

**Supplementary Figure 2** | Transport measurements of a pure $Sb_2Te_3$(25 QL) thin film. **a** Temperature dependent resistivity of this film shows a metallic behavior. Insert shows a schematic drawing of the Hall bar structure and experimental geometry. **b** Hall resistance versus magnetic field. The carrier concentration of this thin film is derived from the slope of the solid fitted line.

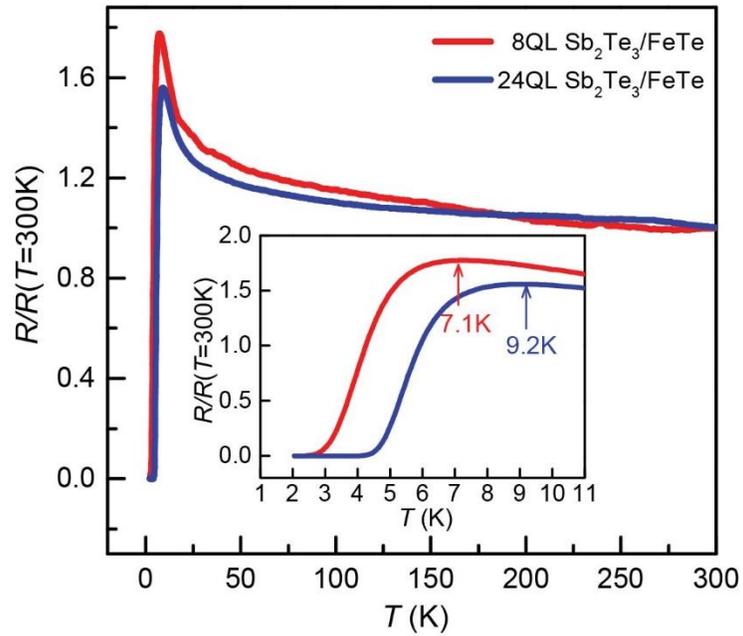

**Supplementary Figure 3** | Normalized resistance curves of $Sb_2Te_3$(8 QL)/$Fe_{1+y}Te$ and $Sb_2Te_3$(24 QL)/$Fe_{1+y}Te$ heterostructures with the $Sb_2Te_3$ layers grown on a single $Fe_{1+y}Te$ layer. The insert shows the curves at low temperature for comparing the superconductivity quality.

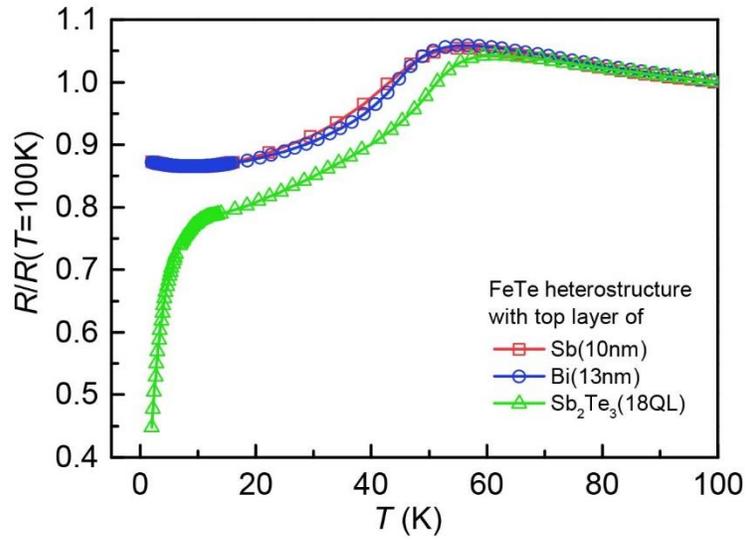

**Supplementary Figure 4 |** Normalized resistance as a function of temperature for three heterostructures, Sb(10 nm)/Fe$_{1+y}$Te, Bi(13 nm)/Fe$_{1+y}$Te and Sb$_2$Te$_3$(18 QL)/Fe$_{1+y}$Te. The antiferromagnetic transition temperatures of these three heterostructures are 54.8, 55.8 and 61.3 K, respectively. Only Sb$_2$Te$_3$(18 QL)/Fe$_{1+y}$Te displays a superconducting transition.

## Supplementary note 1: HRXRD characterization of a $Sb_2Te_3/Fe_{1+y}Te$ heterostructure

The upper part of Supplementary Figure 1 plots the high resolution X-ray diffraction (HRXRD) profile of a $Sb_2Te_3$(27 QL)/$Fe_{1+y}Te$(60 nm) heterostructure. The measurement was carried out in symmetric 2θ-ω scan mode using an x-ray beam (λ = 1.540598 Å) generated from Cu K-α1. The Lower part of Supplementary Figure 1 shows the powder diffraction files of the three crystalline materials contained in this sample, where only the peaks oriented along the normal of the sample surface are extracted for comparison. As can be seen in Supplementary Figure 1, the two strongest peaks are the (002) and (004) peaks of the GaAs substrate, the two second strongest peaks come from the (002) and (004) peaks of the ZnSe(80 nm) buffer layer. Four characteristic diffraction peaks locating at 14.19°, 28.56°, 43.42° and 59.10° match with $Fe_{1+y}Te$ (001), (002), (003) and (004), respectively. From these four peaks, the *c*-axis lattice parameter of $Fe_{1+y}Te$ is determined to be 6.25 Å, which agrees well with the value obtained by STEM as mentioned in the main text. (006), (009), (00 15), (00 18) and (00 21) peaks of $Sb_2Te_3$ can be observed in Supplementary Figure 1 as well, supporting that the $Sb_2Te_3$ layer was likely grown along the [001] direction (note that the 3-index notation is commonly used in HRXRD, which is equivalent to the 4-index notation used in our TEM analysis presented in Fig. 1 in the main text), consistent with our STEM imaging analysis. The Te(~8 nm) capping layer also contributes two detectable peaks.

## Supplementary note 2: Transport results of a pure $Sb_2Te_3$ thin film

A 25 QL-thick pure $Sb_2Te_3$ thin film was grown on a semi-insulating GaAs substrate for resistivity and Hall measurements. Supplementary Figure 2a displays the temperature dependent resistivity curve of this sample, which exhibits a metallic behavior rather than the behavior of an intrinsic topological insulator due to various intrinsic defects. The used Hall bar structure is shown in the same figure with $w = 6$ mm and $l = 1$ mm. No superconductivity (SC) is observed for this sample even the temperature is as low as 2 K. These results together with the fact that pure $Fe_{1+y}Te$ thin films are non-superconducting as shown in Fig. 4a in the main text, indicate that the observed two-dimensional (2D) SC of the $Sb_2Te_3/Fe_{1+y}Te$ heterostructure originates from the interface of the heterostructure. The carrier type in the pure $Sb_2Te_3$ thin film was determined by measuring the magnetic field dependent transverse resistance $R_{xy}$. Supplementary Figure 2b displays $R_{xy}$ as a function of magnetic field $B$ applied perpendicular to the sample surface at 2K and the solid line was fitted with $R_{xy} = \frac{B}{pte}$, where $p$ is the carrier concentration, $t$ the thickness of the thin film and $e$ the electron charge. The positive slope of the fitted solid line indicates that hole-type conductivity dominates in the pure $Sb_2Te_3$ thin film and the hole concentration is determined to be $1.63 \times 10^{22}$ cm$^{-3}$ (It should be noted that this value can only be treated as an estimate of the bulk concentration since such a Hall effect data analysis includes contributions from both surface and bulk carriers). For such a highly degenerate p-type material, its mobility at 2 K can be estimated to be $\mu = \frac{1}{ne\rho} = 3.81$ cm$^2$V$^{-1}$s$^{-1}$.

**Supplementary note 3: Further confirmation of $Sb_2Te_3$ dependence on the 2D SC**

Two heterostructures with the top $Sb_2Te_3$ layer with 8 QL and 24 QL deposited on one single $Fe_{1+y}Te$ layer were fabricated to eliminate the influences on SC caused by the different strength of spin fluctuation, aiming to confirm the TI thickness dependence. The growth of these two samples was achieved with the use of a special sample block having a movable Tantalum clip. First, the whole sample surface was exposed for the growth of a uniform $Fe_{1+y}Te$ layer followed by deposition of 8 QL $Sb_2Te_3$. Then, the Tantalum clip was *in situ* moved to cover one half of the sample surface and a 16 QL-thick $Sb_2Te_3$ layer was deposited on the uncovered half surface. The temperature dependent normalized resistance *R*/*R*(*T*=300 K) curves of these two samples are displayed in Supplementary Figure 3. Neither of them shows a detectable AFM transition, which is attributed to their high spin fluctuation. Insert of Supplementary Figure 3 shows these curves at the temperature near the superconducting transition, which clearly shows that both samples are superconducting with a zero-resistance signature. The onset transition temperature of $Sb_2Te_3$(24 QL)/$Fe_{1+y}$Te is 9.2 K, which is 2.1 K higher than that of $Sb_2Te_3$(8 QL)/$Fe_{1+y}$Te. These results provide further confirmation on the $Sb_2Te_3$ thickness dependence of the quality of the SC as observed in Fig. 5 of the main text.

**Supplementary note 4: SOC alone cannot induce the observed SC**

Three heterostructures, Sb(10 nm)/$Fe_{1+y}$Te, Bi(13 nm)/$Fe_{1+y}$Te and $Sb_2Te_3$(18 QL)/$Fe_{1+y}$Te, were fabricated to test the hypothesis that if the strong spin-orbit coupling (SOC) alone can induce the 2D SC (see main text for detail). Supplementary Figure 4 displays the normalized resistance vs temperature curves of these three heterostructures with their antiferromagnetic (AFM) transition temperatures of 54.8, 55.8 and 61.3 K, respectively, indicating that the fluctuation of the AFM order of the $Sb_2Te_3$/$Fe_{1+y}$Te heterostructure is weaker than that of the Sb/$Fe_{1+y}$Te and Bi/$Fe_{1+y}$Te heterostructures. The fact that the Sb(10 nm)/$Fe_{1+y}$Te and Bi(13 nm)/$Fe_{1+y}$Te heterostructures are non-superconducting indicates that SOC alone cannot induce the interfacial SC as observed in the topological insulator/$Fe_{1+y}$Te heterostructures.